\documentclass[aps, prd, nofootinbib, twocolumn, preprintnumbers, showpacs, floatfix]{revtex4}
\usepackage{graphicx}
\usepackage{amsmath}
\usepackage{multirow}
\usepackage{bm}
\usepackage{color}
\def\fsl#1{\setbox0=\hbox{$#1$}                 
   \dimen0=\wd0                                 
   \setbox1=\hbox{/} \dimen1=\wd1               
   \ifdim\dimen0>\dimen1                        
      \rlap{\hbox to \dimen0{\hfil/\hfil}}      
      #1                                        
   \else                                        
      \rlap{\hbox to \dimen1{\hfil$#1$\hfil}}   
      /                                         
      \fi}                                      %

\newcommand{\VEV}[1]{\langle #1 \rangle}

\begin{document}
\title{Scalar Decay Constant and Yukawa Coupling in Walking Gauge Theories}
\author{Michio Hashimoto}
 \email{michioh@isc.chubu.ac.jp}
  \affiliation{
   Chubu University, \\
   1200 Matsumoto-cho, Kasugai-shi, \\
   Aichi, 487-8501, JAPAN}
\date{\today}

\begin{abstract}
We propose an approach for the calculation of 
the yukawa coupling through the scalar decay constant and 
the chiral condensate in the context of the extended technicolor (ETC).
We perform the nonperturbative computation of the yukawa coupling
based on the improved ladder Schwinger-Dyson equation.
It turns out that the yukawa coupling can be larger or smaller
than the standard model (SM) value,
depending on the number $N_D$ of the weak doublets 
for each technicolor (TC) index.
It is thus nontrivial whether or not 
the huge enhancement of the production of the scalar via the gluon fusion
takes place even for a walking TC model with a colored techni-fermion. 
For the typical one-family TC model near conformality,
it is found that the yukawa coupling is slightly larger than
the SM one, where the expected mass of the scalar bound state 
is around 500~GeV.
In this case, the production cross section via the gluon fusion is 
considerably enhanced, as naively expected, and hence
such a scalar can be discovered/excluded at the early stage of the LHC. 
\end{abstract}

\maketitle

\section{Introduction}

The direct searches for the standard model (SM) Higgs boson have been
intensively performed at the Tevatron~\cite{Aaltonen:2011gs} and 
at the LHC~\cite{Chatrchyan:2011tz,ATLAS-Higgs-search}.
For these Higgs searches, the significant Higgs production process
is the gluon fusion channel. 
If there are extra colored chiral fermions
like in the fourth generation model~\cite{He:2001tp,Frampton:1999xi},
the Higgs production should be enhanced and thus
even a relatively heavy Higgs boson can be surveyed at the early stage of 
the LHC~\cite{ATLAS-higgs-sensitivity,CMS-higgs-sensitivity}. 

The walking technicolor (WTC) is a candidate of 
the dynamical electroweak symmetry breaking (DEWSB) scenario~\cite{Holdom:1981rm,Yamawaki:1985zg,Akiba:1985rr,Appelquist86}.
It can resolve the problems of the flavor changing neutral current (FCNC), 
too light fermion masses and too light pseudo Nambu-Goldstone (NG) bosons,
which were serious difficulties in 
the QCD-like TC~\cite{miransky-textbook,Hill:2002ap}. 
Although the QCD-like TC was strongly disfavored by
the precision measurements~\cite{Peskin-Takeuchi},
the estimate of the $S$-parameter in the QCD-like TC is 
not applicable for the WTC.
Evidence of the reduction of the $S$-parameter is reported
in the ladder Schwinger-Dyson (SD) and Bethe-Salpeter (BS) 
approach~\cite{Harada:2005ru}, and 
also in the lattice simulation~\cite{Appelquist:2010xv}.
In the holographic WTC model, one can find a parameter space with 
a small $S \; (\sim 0.1)$~\cite{Hong:2006si,Haba:2008nz}.

A ``light'' scalar, so-called the techni-dilaton (TD),
which is the pseudo NG boson associated with the scale symmetry breaking,
is predicted in the WTC~\cite{Yamawaki:1985zg,Bando:1986bg,Holdom:1986ub}.
The TD mass near the critical point has been suggested as 
$M_{\rm TD} \sim \sqrt{2}m$ in the context of 
the gauged Nambu-Jona-Lasinio (NJL) model~\cite{Shuto:1989te}, 
where $m$ represents the dynamically generated fermion mass.
For recent discussions on the TD mass in the criticality limit, 
see Refs.~\cite{Hashimoto:2010nw,Dietrich:2005jn}.
The straightforward calculation in the ladder SD and BS approach
suggests numerically $M_{\rm TD} \sim 500$~GeV 
for the typical one-family TC model~\cite{Harada:2003dc}.

It is noticeable that the early stage of the LHC has the sensitivity
to such a heavy Higgs~\cite{ATLAS-higgs-sensitivity,CMS-higgs-sensitivity}. 
Notice that the gluon fusion process counts 
the number of the colored particles and 
also depends on the magnitude of their yukawa couplings.
In particular, the estimate of the yukawa coupling is not so trivial
in the DEWSB scenario, 
because a nonperturbative computation is inevitably required.

In this paper, we propose an approach for the calculation of 
the yukawa coupling through the scalar decay constant and 
the chiral condensate.
We will adopt the ladder SD approach as a nonperturbative method.
In principle, these values would be extracted from the lattice simulation.

Let us derive a relation between the scalar decay constant and 
the yukawa coupling.

\begin{figure}[t]
  \begin{flushleft}
  \hspace*{2cm}
  \resizebox{0.1\textwidth}{!}{\includegraphics{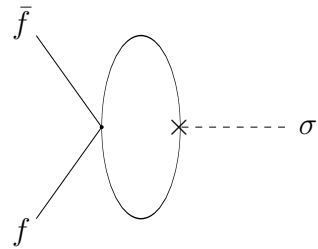}}
  \end{flushleft}
  \caption{Yukawa coupling between the SM fermions $f$ and 
   the scalar bound state $\sigma$ in the framework of the ETC.
   The techni-fermion loop generates the mass of $f$ and 
   also intermediates between $f$ and $\sigma$.
  \label{fig-yukawa}}
\end{figure}

Suppose that the extended technicolor (ETC) sector generates
the four-fermion interaction,
\begin{equation}
  {\cal L}_{4F} = G_f \bar{\psi}\psi \bar{f}f,
  \label{4F}
\end{equation}
where $\psi$ and $f$ denote the techni and SM fermions.
The SM fermion mass $m_f$ is obtained from the techni-fermion condensate,
\begin{equation}
    m_f = - G_f Z_m^{-1} \VEV{\bar{\psi}\psi}_R , 
    \label{mf}
\end{equation}
with the renormalization constant $Z_m \sim m/\Lambda_{\rm ETC}$,
where $\Lambda_{\rm ETC}$ is the ETC scale and
the subscript $R$ represents the renormalized quantity.
The scalar decay constant $F_\sigma$ for the scalar current is defined by
\begin{equation}
 \langle 0|(\bar{\psi}\psi(0))_R| \sigma (q) \rangle \equiv 
 F_\sigma M_\sigma , 
 \label{def-Fs}
\end{equation}
where $M_\sigma$ is the mass of the scalar bound state $\sigma$. 
Eqs.~(\ref{mf}) and (\ref{def-Fs}) immediately yield 
the following expression of the yukawa coupling between $\sigma$ and $f$, 
\begin{equation}
  g_{\sigma ff} = Z_m^{-1} G_f F_\sigma M_\sigma =
  \frac{m_f}{\frac{-\VEV{\bar{\psi}\psi}_R}{F_\sigma M_\sigma}}\,.
  \label{yukawa-Ftd}
\end{equation}
For a graphical expression, see Fig.~\ref{fig-yukawa}.
Since the SM yukawa coupling is given by
$g_{hff}^{\rm SM} = m_f/v$ with $v = 246$~GeV, 
the ratio of the two is
\begin{equation}
  \frac{g_{\sigma ff}}{g_{hff}^{\rm SM}} =
  \frac{v}{\frac{-\VEV{\bar{\psi}\psi}_R}{F_\sigma M_\sigma}}  \, .
  \label{ratio-y}
\end{equation}
We can calculate $F_\sigma$ through the correlation function $\Pi_\sigma$ 
for the scalar operator, which is defined by
\begin{equation}
  {\cal F.T.} 
  i\langle 0|(\bar{\psi}\psi(x))_R (\bar{\psi}\psi(0))_R | 0 \rangle 
  \equiv \Pi_\sigma (q) \, .
\end{equation}
The scalar mass $M_\sigma$ and the scalar decay constant $F_\sigma$ 
can be read from the pole and residue of $\Pi_\sigma(q)$, 
owing to the spectral representation,
\begin{equation}
  \Pi_\sigma(q) = \frac{F_\sigma^2 M_\sigma^2}{-q^2 + M_\sigma^2} \, .
\end{equation}
Note that $\Pi_\sigma(0)=F_\sigma^2$ in this normalization.
On the other hand, the (renormalized) second derivative of 
the effective potential at the stationary point corresponds to 
the inverse of the two point function at the zero momentum,
\begin{equation}
  \frac{d^2 V}{d\sigma_R^2} = \Pi_\sigma^{-1}(0) = \frac{1}{F_\sigma^2}, 
\end{equation}
and thereby it holds
\begin{equation}
  \sigma_R^2 \frac{d^2 V}{d\sigma_R^2} = 
  \left(\frac{-\VEV{\bar{\psi}\psi}_R}{F_\sigma}\right)^2,
\end{equation}
which is closely connected with the yukawa coupling 
via Eq.~(\ref{yukawa-Ftd}).
We emphasize that this quantity is obviously independent of 
the renormalization point.

We perform the calculations of $F_\sigma$ and $\VEV{\bar{\psi}\psi}_R$
by using the improved ladder SD equation~\cite{Hashimoto:2010nw}. 
For a given $M_\sigma$, the yukawa coupling is estimated from 
Eq.~(\ref{yukawa-Ftd}).
We then find the ratio of the yukawa coupling 
$g_{\sigma ff}/g_{hff}^{\rm SM} \simeq 1.2$ 
for the typical one-family TC model with $N_{\rm TC}=2$ and 
$M_{\sigma} = 500$~GeV under the realistic setup 
$m/\Lambda_{\rm ETC} \sim 10^{-3}\mbox{--}10^{-4}$.
The yukawa coupling was estimated also by using the Ward-Takahashi (WT) 
identity and a hypothesis of the partially conserved dilaton current 
(PCDC)~\cite{Bando:1986bg}.
The numerical result via the scalar decay constant agrees with 
the PCDC approach~\cite{Hashimoto:2010nw}.

In the one-family TC model, the colored techni-fermions
contribute to the production of $\sigma$, furthermore.
The cross section is thus considerably enhanced.
Such a model should be confirmed/excluded at the early stage of the LHC.
On the other hand, it is not the case 
for the model having only one weak doublet and no techni-quark. 

This paper is organized as follows:
In Sec.~\ref{sec2}, we study the improved ladder SD equation.
An analytical expression for the mass function is derived.
In Sec.~\ref{sec3}, we first review the formalism of the effective potential.
By using the analytical expression of the mass function,
we calculate $F_\sigma$ and $\VEV{\bar{\psi}\psi}_R$.
We then obtain the yukawa coupling $g_{\sigma ff}$.
Also, the phenomenological implications are briefly discussed. 
Sec.~\ref{sec4} is devoted to summary and discussions.

\section{Gap equation}
\label{sec2}

We adopt the ladder SD equation as a nonperturbative approach.
In order to incorporate the running effects of the gauge coupling,
the improved one has been studied~\cite{Miransky:1983vj}.
With the bare mass $m_0$, it is written by
\begin{equation}
  B(x) = m_0 + \int_0^{\Lambda^2} dy
  \frac{y B(y)}{y+B^2(y)}\frac{\lambda(\max(x,y))}{\max(x,y)},
  \label{imp-ladder}
\end{equation}
where $x$ and $y$ represent the Euclidean momenta, and
the normalized gauge coupling $\lambda(x)$ is defined by 
\begin{equation}
  \lambda(x) \equiv \frac{3C_F \, \alpha(\mu^2=x)}{4\pi} \, .
\end{equation}
We also introduced the cutoff $\Lambda$ for the ladder SD equation. 

In the two-loop approximation,
the renormalization group equation (RGE) of $\alpha$ is~\cite{pdg}
\begin{equation}
  \mu^2 \frac{\partial }{\partial \mu^2}\alpha = \beta (\alpha)=
  -b_0 \alpha^2 -b_1 \alpha^3, 
  \label{banks-zaks}
\end{equation}
with
\begin{equation}
  b_0 = \frac{1}{12\pi}(11C_A - 4 N_f T_R), 
\end{equation}
and
\begin{equation}
    b_1 = \frac{1}{24\pi^2}\bigg[\,17C_A^2 - 2 N_f T_R(5C_A+3C_F)\,\bigg], 
\end{equation}
where $N_f$ represents the number of flavor and
the group theoretical factors are 
\begin{equation}
  C_A=N_{\rm TC}, \quad T_R = \frac{1}{2}, \quad 
  C_F = \frac{N_{\rm TC}^2-1}{2N_{\rm TC}},
\end{equation}
for $SU(N_{\rm  TC})$ gauge theories.

When $b_0 > 0$ and $b_1 < 0$, the Caswell--Banks--Zaks 
infrared fixed point (CBZ-IRFP)~\cite{Caswell:1974gg} emerges, 
\begin{equation}
  \alpha_* \equiv \frac{b_0}{-b_1} \, .
\end{equation}
By using the CBZ-IRFP $\alpha_*$ and 
the Lambert function $W$~\cite{Corless:1996zz}, 
which is the inverse of $x e^x$,
we can express $\alpha(x)$ analytically~\cite{Gardi:1998rf},
\begin{equation}
  \alpha (x) = \frac{\alpha_*}{1+W(z(x))},
  \label{alp-lambert}
\end{equation}
where $z$ is defined by
\begin{equation}
  z(x) \equiv
  \frac{1}{e}\left(\frac{x}{\Lambda_{\rm I}^2}\right)^{b_0 \alpha_*},
  \label{z-lam0}
\end{equation}
with the intrinsic scale $\Lambda_{\rm I} \; (\sim \Lambda_{\rm ETC})$, 
which is analogous to the QCD scale $\Lambda_{\rm QCD}$.

The integral form (\ref{imp-ladder}) can be rewritten by 
the differential equation with the IR and UV boundary conditions (BC's). 
Ignoring $x\frac{d \lambda}{dx} (\propto \beta \ll \lambda)$,
the differential form is
\begin{equation}
     x^2 \frac{d^2}{dx^2} B(x) 
  + 2x  \frac{d}{dx} B(x)
  + \lambda (x) \frac{x B(x)}{x+B^2(x)} = 0,
 \label{sde-diff}
\end{equation}
and the two BC's are
\begin{eqnarray}
\mbox{(UV-BC)}: 
&&  x \frac{d}{dx} B(x)\Bigg|_{x=\Lambda^2} + B(\Lambda^2)=m_0, 
    \label{uvbc} \\
\mbox{(IR-BC)}: 
&& x^2 \frac{d}{dx} B(x) \Bigg|_{x \to 0} \to 0 \, .
   \label{irbc}
\end{eqnarray}

Let us solve analytically the improved ladder SD equation (\ref{sde-diff})
in the following approximations.

By using the bifurcation method and also 
the parabolic deformation of the RGE~\cite{Hashimoto:2010nw},
\begin{equation}
  \beta (\alpha) \to -b_0 \alpha (\alpha_* - \alpha) , 
\end{equation}
whose solution is
\begin{equation}
  \alpha(x) = 
  \frac{\alpha_*}
       {1 + e^{-1}\left(\frac{x}{\Lambda_{\rm I}^2}\right)^{b_0 \alpha_*}} , 
  \label{alp-parabolic}
\end{equation}
we analytically obtain the solution of the linearized ladder SD equation,
\begin{widetext}
\begin{eqnarray}
  \frac{B(x)}{B_0} &=& \phantom{+}
  c_1 \left(\frac{x}{B_0^2}\right)^{-\frac{1-\omega}{2}} 
  F\bigg(-\frac{1-\omega}{2s},\frac{1+\omega}{2s},1+\frac{\omega}{s};
         1-\frac{\lambda_*}{\lambda(x)}\bigg)
  \nonumber \\
&&
 +d_1 \left(\frac{x}{B_0^2}\right)^{-\frac{1+\omega}{2}} 
  F\bigg(-\frac{1+\omega}{2s},\frac{1-\omega}{2s},1-\frac{\omega}{s};
         1-\frac{\lambda_*}{\lambda(x)}\bigg) , \qquad (x \geq B_0^2),
  \label{parabolic-sol}
\end{eqnarray}
\end{widetext}
where $F(\alpha,\beta,\gamma;z)$ represents 
the Gauss's hypergeometric function and
$B_0$ is the normalization factor of the mass function
defined by $B(x=B_0^2)=B_0$.
We also introduced the normalized IR-FP $\lambda_*$, 
\begin{equation}
  \lambda_* \equiv \frac{3C_F \alpha_*}{4\pi} , 
\end{equation}
the power factor $s$,
\begin{equation}
  s \equiv b_0 \alpha_* > 0 , 
\end{equation}
and 
\begin{equation}
  \omega \equiv \sqrt{1-\frac{\lambda_*}{\lambda_{\rm cr}}} , \qquad
  \lambda_{\rm cr} \equiv \frac{1}{4} \, .
\end{equation}
The integration constants $c_1$ and $d_1$ are
determined through the IR-BC and the normalization of $B(x)$.
In the limit of $B_0 \ll \Lambda_{\rm I}$, we obtain
\begin{equation}
  c_1 = \frac{1+\omega}{2\omega}, \qquad
  d_1 = -\frac{1-\omega}{2\omega} \, .  
\end{equation}
The UV-BC yields a relation among $m_0$, $B_0$, $\Lambda_{\rm I}$
and $\omega$.
In the chiral limit $m_0 \to 0$,
it turns out that there appears a nontrivial solution $B_0 \to m \ne 0$,
only when $\omega$ is pure imaginary $\omega = i\tilde{\omega}$, 
\begin{equation}
 \tilde{\omega} \equiv \sqrt{\frac{\lambda_*}{\lambda_{\rm cr}}-1}>0, \quad
 \mbox{i.e.,} \quad \lambda_* > \lambda_{\rm cr} \, .
\end{equation}
This approximation {\it qualitatively} works well.
For details, see Ref.~\cite{Hashimoto:2010nw}.

Another linearizing method is to replace the denominator of
the last term of Eq.~(\ref{sde-diff}) by $x+B_0^2$~\cite{miransky-textbook}.
Introducing $\xi \equiv (B_0/\Lambda_{\rm I})^{2s}$ and 
$B(x)/B_0 \equiv \Sigma^{(0)}(x) + \xi \Sigma^{(1)}(x)+\cdots$,
and also expanding $\lambda(x)$ by $\xi$,
we can solve the linearized ladder SD equation 
owing to the analytic form of $\Sigma^{(0)}(x)$,
$\Sigma^{(0)}(x)=F((1+\omega)/2,(1-\omega)/2,2;-x^2/B_0^2)$~\cite{miransky-textbook}. 
The solution in the region $x \gg B_0^2$ and 
$|\lambda(x)/\lambda_* -1| \ll 1$ is similar to
the bifurcation solution (\ref{parabolic-sol}).

For both linearized solutions, 
we find the behavior of the mass function 
in the region where the momentum is large in the sense that $x \gg B_0^2$
and the gauge coupling is slowly running, $\lambda(x)/\lambda_* \approx 1$, 
\begin{equation}
  \frac{B(x)}{B_0} \simeq \frac{A}{\tilde{\omega}}
  \left(\frac{x}{B_0^2}\right)^{-\frac{1}{2}}
  \sin \left(\frac{\tilde{\omega}}{2}\ln \frac{x}{B_0^2} + \delta\right),
  \label{B-app}
\end{equation}
where $A$ and $\delta$ are
\begin{equation}
  A = \sqrt{1+\tilde{\omega}^2}, \quad \delta = \arctan \tilde{\omega},
\end{equation}
for the former approximation, and 
\begin{equation}
  A = 2|C|, \quad
  e^{2i\delta} = \frac{C}{C^*} , \quad
  C \equiv \frac{\Gamma(1+i\tilde{\omega})}
                {\Gamma\left(\frac{1+i\tilde{\omega}}{2}\right)
                 \Gamma\left(\frac{3+i\tilde{\omega}}{2}\right)}, 
\end{equation}
for the latter one.
In particular, for the former and the latter,
$A \to 1$ and $A \to 4/\pi$ in the limit $\tilde{\omega} \to 0$,
respectively.
It means that the analytic property of the mass function 
can be {\it qualitatively} approximated by Eq.~(\ref{B-app}),
while there are {\it quantitatively} ambiguities about 
the values of $A$ and $\delta$. 
In the next section, we will fix this quantitative uncertainty 
by using the numerical analysis of the ladder SD equation 
with the two-loop running coupling~\cite{Hashimoto:2010nw}.

\section{Estimate of the yukawa coupling}
\label{sec3}

In order to calculate $F_\sigma$ and $g_{\sigma ff}$, 
we study the effective potential for $\sigma$.

Following Ref.~\cite{Miransky:1992bj,Gusynin:2002cu},
we review the formalism on the effective potential. 

The generating functional is defined by
\begin{equation}
  W[J] \equiv \frac{1}{i} \ln \int [d\psi d\bar{\psi}][\mbox{gauge}]
  e^{i\int d^4x ({\cal L} + J \bar{\psi}\psi)} \, .
\end{equation}
and also the effective action is 
\begin{equation}
  \Gamma[\sigma] \equiv W[J] - \int d^4 x J \sigma , 
\end{equation}
where
\begin{equation}
  \sigma(x) \equiv \bar{\psi}(x) \psi(x) \, . 
\end{equation}
For constants $\sigma$ and $J$,
the effective potential is defined by 
$V = -\Gamma[\sigma]/\int d^4x$.
Noting that
\begin{equation}
  \frac{d V(\sigma)}{d \sigma} = J,
\end{equation}
the effective potential is formally given by
\begin{equation}
  V(\sigma) = \int d\sigma J , 
\end{equation}
where $J$ should be regarded as a function of $\sigma$
in this expression. 

In the context of the ladder SD equation,
it is convenient to use the IR mass $B_0$ 
in the expressions of $\sigma$ and $J$.
We then obtain
\begin{equation}
  V(\sigma) = \int dB_0 \frac{d\sigma(B_0)}{dB_0} J(B_0) , 
\end{equation}
where $B_0$ should be transformed into $\sigma$ after the integral.
Also, the second derivative of the effective potential is 
\begin{equation}
  \frac{d^2 V}{d\sigma^2} =
  \frac{d J}{dB_0} \left(\frac{d\sigma}{dB_0}\right)^{-1} \, .
\end{equation}

Let us explicitly calculate the effective potential
and its second derivative.
In the following calculation, it is enough to take
the cutoff $\Lambda$ in the ladder SD equation~(\ref{imp-ladder})
to the ETC scale,
\begin{equation}
  \Lambda \to \Lambda_{\rm ETC} \, .
\end{equation}
We consider only the case with $\lambda_* > \lambda_{\rm cr}$.

The effect of the constant source $J$ is obtained by 
the replacement 
\begin{equation}
  m_0 \to m_0 - J \, .
\end{equation}
Note that the UV-BC (\ref{uvbc}) yields
\begin{equation}
  \Lambda_{\rm ETC}^2 B'(\Lambda_{\rm ETC}^2) + B(\Lambda_{\rm ETC}^2) = m_0 - J,
\end{equation}
where $B'(x) \equiv \frac{dB(x)}{dx}$. 
The bare chiral condensation is 
\begin{equation}
  \sigma \equiv \VEV{\bar{\psi}\psi} = 
  - \frac{N_{\rm TC} N_f}{4\pi^2}
    \int_0^{\Lambda_{\rm ETC}^2} dx x \frac{B(x)}{x+B^2(x)} \, .
\end{equation}
By using the ladder SD equation (\ref{imp-ladder}), we also find
\begin{equation}
  \sigma = \frac{N_{\rm TC} N_f}{4\pi^2}
  \frac{\Lambda_{\rm ETC}^4}{\lambda_{\rm ETC}} 
  B'(\Lambda_{\rm ETC}^2) , 
\end{equation}
where $\lambda_{\rm ETC} \equiv \lambda(\Lambda_{\rm ETC}^2) \sim \lambda_*$ 
and we ignored $x\frac{d\lambda(x)}{dx} \propto \beta \ll \lambda$.

We may employ the approximation (\ref{B-app}) in the UV region,
and hence obtain
\begin{eqnarray}
&&  \hspace*{-4mm}
   m_0 - J = \nonumber \\
&&\frac{A\sqrt{1+\tilde{\omega}^2}}{2\tilde{\omega}}
  \frac{B_0^2}{\Lambda_{\rm ETC}} 
  \sin\Bigg(\tilde{\omega}\ln \frac{\Lambda_{\rm ETC}}{B_0} + 
  \delta + \arctan \tilde{\omega}\Bigg),\nonumber \\
  \label{sde-m0-j}
\end{eqnarray}
and 
\begin{align}
  \sigma &= - \frac{N_{\rm TC} N_f}{4\pi^2}
  \frac{B_0^2 \Lambda_{\rm ETC}}{\lambda_{\rm ETC}} \nonumber \\
& \quad 
  \frac{A\sqrt{1+\tilde{\omega}^2}}{2\tilde{\omega}} 
  \sin\Bigg(\tilde{\omega}\ln \frac{\Lambda_{\rm ETC}}{B_0} + 
  \delta - \arctan \tilde{\omega}\Bigg) \, .
\end{align}
The stationary condition $\frac{d V}{d\sigma}=J=0$
in the chiral limit $m_0 \to 0$ gives 
the solution of the ladder SD equation, 
$\tilde{\omega}\ln \Lambda_{\rm ETC}/B_0 + \delta + 
\arctan \tilde{\omega} = n \pi$, $(n=1,2,3,\cdots)$.
It is known that the zero node solution $B_0 = B_0^{(1)} \equiv m$ 
corresponds to the true vacuum~\cite{miransky-textbook}.

We renormalize $\sigma$ with fixing the zero node solution $m$.
The renormalized quantity at the true vacuum $B_0 = m$ is 
\begin{equation}
  \sigma_R = Z_m \sigma \to 
  \VEV{\bar{\psi}\psi}_R = - \frac{N_{\rm TC} N_f}{4\pi^2}
   \frac{A}{\lambda_* \sqrt{1+\tilde{\omega}^2}}\, m^3 ,
\end{equation}
where $Z_m \sim m/\Lambda_{\rm ETC}$ and 
$\lambda_{\rm ETC}$ is also renormalized to $\lambda_*$.
It is straightforward to calculate the vacuum energy, i.e., 
the value of the effective potential at the true vacuum,
\begin{equation}
  V_{\rm sol} = V|_{B_0 = m} =
 - \frac{N_{\rm TC} N_f}{4\pi^2} \frac{A^2}{16 \lambda_*}\, m^4 \, .
 \label{V}
\end{equation}
Note that in the limit of $\tilde{\omega} \to 0$ 
($\lambda_* \to \lambda_{\rm cr}=1/4$), 
Eq.~(\ref{V}) with $A \to 4/\pi$ reproduces 
the expression of the vacuum energy in Ref.~\cite{Miransky:1989qc}.
We also find the renormalized second derivative of the effective potential 
at the true vacuum, which corresponds to the inverse of 
the square of the scalar decay constant, 
\begin{equation}
  \frac{1}{F_\sigma^2} = \frac{d^2 V}{d\sigma_R^2}\Bigg|_{B_0 \to m}  
  = \frac{1+\tilde{\omega}^2}{\frac{N_{\rm TC} N_f}{4\pi^2}(5-\tilde{\omega}^2)}
    \frac{\lambda_*}{m^2} \, .
  \label{fsigma}
\end{equation}
The square of the chiral condensate over the scalar decay constant is then
\begin{equation}
  \left(\frac{-\VEV{\bar{\psi}\psi}_R}{F_\sigma}\right)^2
  = \sigma_R^2 \frac{d^2 V}{d\sigma_R^2}
  = \frac{N_{\rm TC} N_f}{4\pi^2}\frac{A^2}{\lambda_*}
    \frac{1}{5-\tilde{\omega}^2} \, m^4 \, .
\end{equation}

We determine the value of $A$ so as to reproduce 
the vacuum energy in the numerical analysis of 
the improved ladder SD equation with
the two-loop running coupling~\cite{Hashimoto:2010nw},
\begin{equation}
  \VEV{\theta_\mu^\mu} = 4V_{\rm sol} \equiv 
 -\frac{N_{\rm TC} N_f}{2\pi^2} \kappa_V^{} m^4 \, .
\end{equation}
We show the numerical values of $\kappa_V^{}$ and 
the determined $A$ in Table~\ref{tab1}.

\begin{table*}
\begin{equation*}
  \begin{array}{c|c|c|c||c||c||c}\hline
    \lambda_* & \frac{m}{\Lambda_{\rm ETC}} & \kappa_V^{} & \kappa_F^{} &
     A & \sqrt{\frac{N_D}{N_f}} \frac{F_\sigma}{v} & 
     \frac{g_{\sigma ff}}{g_{h ff}^{\rm SM}} \frac{v}{N_D M_\sigma} \\ \hline
   0.305 & 1.12 \times 10^{-3} & 0.685 & 1.38 & 1.29 & 2.59 & 0.142 \\
   0.287 & 1.08 \times 10^{-4} & 0.709 & 1.42 & 1.28 & 2.71 & 0.148 \\
   0.258 & 5.88 \times 10^{-10} & 0.756 & 1.48 & 1.25 & 2.93 & 0.157 \\ \hline
 \end{array}
\end{equation*}
\caption{Estimates of $A$, $F_\sigma/v$ and $g_{\sigma ff}/g_{hff}^{\rm SM}$ 
        for several values of $\lambda_*$. 
        We read the corresponding values of $m/\Lambda_{\rm ETC}$,
        $\kappa_V^{}$ and $\kappa_F$ from the numerical analysis 
        of the ladder SD equation with the two-loop running 
        coupling~\cite{Hashimoto:2010nw}.
         \label{tab1}}
\end{table*}

\begin{table*}
  \begin{equation*}
  \begin{array}{l||c||c|c|c|c}\hline
    & \frac{\sigma (gg \to \sigma)}{\sigma (gg \to h_{\rm SM})} &
      \mbox{Br}(\sigma \to WW) & \mbox{Br}(\sigma \to ZZ) &
    \mbox{Br}(\sigma \to t\bar{t}) & \mbox{Br}(\sigma \to gg) \\ \hline
    \mbox{One-Family model } (N_{\rm TC}=2) & 36 & 30\% & 14\% & 52\% & 
    4\% \\ \hline
    \mbox{Techni-quark } (N_{TQ}=4, N_D=1) & 2.2 & 30\% & 14\% & 52\% & 
    4\% \\ \hline
    \mbox{No Techni-quark } (N_{TQ}=0, N_D=4) & 1.2 & 31\% & 15\% & 54\% & 
    \sim 0\% \\ \hline
    \mbox{No Techni-quark } (N_{TQ}=0, N_D=1) & 0.090 & 31\% & 15\% & 54\% & 
    \sim 0\% \\ \hline
  \end{array}
  \end{equation*}
  \caption{Ratio of the production cross section of $\sigma$ 
   to that of the SM Higgs and the branching ratios. 
   The mass of $\sigma$ is fixed to $M_\sigma=500$~GeV.
   We took $N_{\rm TC}=2$, $N_f=8$ and $m/\Lambda_{\rm ETC}=1.08 \times 10^{-4}$.
   For comparison with the typical one-family TC model, 
   we just varied the numbers $N_{TQ}$ and $N_D$ in the second, third and
   fourth rows.
   \label{tab2}}
\end{table*}

The pseudo-scalar decay constant $F_\pi$ is connected with 
the weak scale.
For the estimate of $F_\pi$, 
it is convenient to employ the Pagels-Stokar formula~\cite{Pagels:1979hd}, 
\begin{equation}
  F_\pi^2 = \frac{N_{\rm TC}}{4\pi^2} \int_0^{\Lambda_{\rm ETC}^2} dx x 
  \frac{B^2(x)-\frac{x}{4}\frac{dB^2(x)}{dx}}{(x+B^2(x))^2}\, .
\end{equation}
The numerical factor $\kappa_F$ between $m$ and $F_\pi$ is defined by
\begin{equation}
  v^2 = N_D F_\pi ^2 \equiv \frac{N_{\rm TC} N_D}{4\pi^2} \, \kappa_F^2\, m^2 ,
  \label{fpi}
\end{equation}
where $N_D$ denotes the number of the weak doublets for each TC index.
By definition, $N_D \leq N_f/2$.
In Table~\ref{tab1}, we show the values of $\kappa_F$ 
calculated in Ref.~\cite{Hashimoto:2010nw}.
Since the normalization of the mass function $B(x=m^2)=m$ yields 
$\delta = \arcsin (\tilde{\omega} A^{-1})$ in the approximation (\ref{B-app}),
we can estimate $F_\pi$ by using Eq.~(\ref{B-app})
with the values of $A$ in Table~\ref{tab1}.
We found that the differences of $\kappa_F$ are about 
2\%, 1\% and 1\% from top to bottom in Table~\ref{tab1}, respectively.
Although the approximation~(\ref{B-app}) is inapplicable in the IR region,
it practically works well.

We now describe $F_\sigma$, $\VEV{\bar{\psi}\psi}_R$
and $g_{\sigma ff}$ more explicitly.
Eqs.~(\ref{fsigma}) and (\ref{fpi}) yield
\begin{equation}
  \frac{F_\sigma}{v} = \sqrt{\frac{N_f}{N_D}}
  \sqrt{\frac{5-\tilde{\omega}^2}{(1+\tilde{\omega}^2)\lambda_*}}
  \frac{1}{\kappa_F} , 
\end{equation}
the renormalized chiral condensate is
\begin{equation}
  \frac{-\VEV{\bar{\psi}\psi}_R}{v^3} = \frac{N_f}{N_D\sqrt{N_{\rm TC}N_D}}
  \frac{4\pi \sqrt{2\kappa_V}}{\kappa_F^3\sqrt{(1+\tilde{\omega}^2)\lambda_*}},
\end{equation}
and hence the ratio of the yukawa coupling (\ref{ratio-y}) reads
\begin{equation}
  \frac{g_{\sigma ff}}{g_{hff}^{\rm SM}} = \sqrt{\frac{N_{\rm TC}}{N_f}} N_D 
  \frac{\kappa_F^2 \sqrt{5-\tilde{\omega}^2}}{4\pi\sqrt{2\kappa_V^{}}}
  \frac{M_\sigma}{v} \, .
\end{equation}
We show the numerical values of $F_\sigma/v$ and 
$g_{\sigma ff}/g_{hff}^{\rm SM}$ in Table~\ref{tab1}.
The yukawa coupling for the techni-fermions should be almost the same.

We here note that $N_f \simeq 4N_{\rm TC}$ in the walking gauge theory.
For the yukawa coupling in Table~\ref{tab1}, we have already put 
$N_f = 4N_{\rm TC}$.
On the other hand, the number $N_D$ is model-dependent;
If all flavors have the weak charge like in the typical one-family TC model, 
$N_D=N_f/2$. The minimum case is $N_D=1$.

The scalar mass $M_\sigma$ is closely related to 
the dynamically generated techni-fermion mass $m$ 
in the ladder SD approach.
The values of $m$ are estimated from Eq.~(\ref{fpi}).
For the typical one-family TC model with $N_{\rm TC}=2$, $N_f=8$
and $N_D=4$, it reads $m=$ 390~GeV, 380~GeV, 370~GeV from top 
to bottom in Table~\ref{tab1}. 
The handy mass formula in the critical limit of 
the gauged NJL model, $M_\sigma \simeq \sqrt{2}m$~\cite{Shuto:1989te}, 
then yields $M_\sigma \simeq$ 560~GeV, 540~GeV, 520~GeV, respectively.
These are consistent with the estimate in the BS approach,
$M_\sigma \sim v \sqrt{17/N_D} \sim$ 500~GeV~\cite{Harada:2003dc}.
For a fixed value of the scalar mass, $M_\sigma = $ 500~GeV,
the ratios of $M_\sigma$ and $m$ are $M_\sigma/m=1.3, 1.3, 1.4$, 
respectively.

In Ref.~\cite{Bando:1986bg}, 
by using the WT identity and the PCDC relation, 
$F_{\rm TD}^2 M_{\rm TD}^2 = -4 \VEV{\theta_\mu^\mu}$,
the yukawa coupling $y_{\rm TD}$ between the TD and the SM fermions 
was estimated as $y_{\rm TD} = (3-\gamma_m) m_f/F_{\rm TD}$.
Note that in general, the scalar decay constant $F_\sigma$ discussed
in this paper is different from the TD decay constant $F_{\rm TD}$,
which is $\langle 0 | \theta_\mu^\mu(0)|TD\rangle = F_{\rm TD} M_{\rm TD}^2$.
The estimate of $v/F_{\rm TD}$ is $3/5$ for the typical one-family TC model
with $N_{\rm TC}=2$ and $M_{\rm TD}=500$~GeV~\cite{Hashimoto:2010nw}.
This yields $y_{\rm TD}/g_{hff}^{\rm SM}=1.2$ and thus numerically agrees with 
the corresponding result of $g_{\sigma ff}/g_{hff}^{\rm SM}$.

Let us briefly discuss the phenomenological implications.

As suggested above, we may identify the scalar bound state $\sigma$ 
to the TD.
We have already shown that the yukawa coupling is different from 
that in the usual dilaton~\cite{Goldberger:2007zk},
when the masses of the SM fermions are originated from 
the four-fermion interactions as in the ETC.
On the other hand, it is reasonable to assume 
the couplings of $\sigma$ and the weak bosons are
$g_{\sigma WW}/g_{hWW}^{\rm SM} = g_{\sigma ZZ}/g_{hZZ}^{\rm SM} = v/F_{\rm TD}$, 
as usual~\cite{Goldberger:2007zk},
where $g_{hWW,hZZ}^{\rm SM}$ are the SM values.

For a typical mass, $M_\sigma=500$~GeV,
we estimate the enhancement factor of the production cross section of 
$\sigma$ via the gluon fusion process and the branching ratios,
assuming that there are no other light resonances like the techni-pion.
The results for the typical one-family TC model and others
are shown in Table~\ref{tab2}. 

For the one-family TC model, which contains $N_{TQ}=2N_{\rm TC}$ extra quarks 
(colored techni-fermions),
the enhancement factor of the production of $\sigma$ 
compared with the SM is huge $\sim 36$ in the heavy quark limit, 
even for $N_{\rm TC}=2$. 
Such a model should be confirmed/excluded at the early stage of the LHC.

Concerning this large enhancement over 10 times, 
$N_{TQ}$ and also $N_D$ are crucial.
As a demonstration, we may just reduce the numbers $N_{TQ}$ and/or $N_D$.
The results are shown in Table~\ref{tab2}. 
For the models with $N_{TQ}=4,N_D=1$ and $N_{TQ}=0,N_D=4$,
the production cross section is fairly comparable to the SM one.
However, for the model having only one weak doublet of the techni-fermion
and no techni-quark, the production cross section gets much smaller. 

It might be worthwhile to mention that compared with the SM,
the branching ratio to the top-pair is increasing and 
that to the weak bosons is decreasing.

We have studied the yukawa coupling in the framework of the ETC.
However the ETC sector might not produce fully 
the top quark mass~\cite{Hill:1994hp,Hill:2002ap}.
In such a class of models, 
the gluon fusion would not be the main production channel. 
This is out of scope in this paper.

\section{Summary and discussions}
\label{sec4}

We proposed the alternative approach to estimate 
the yukawa coupling via the scalar decay constant and 
the chiral condensate. 
By using the improved ladder SD approach, we calculated 
$F_\sigma$, $\VEV{\bar{\psi}\psi}_R$ and $g_{\sigma ff}$.

For the typical one-family TC model with $N_{\rm TC}=2$,
we numerically found $g_{\sigma ff}/g_{hff}^{\rm SM} \simeq 1.2$
under the realistic setup $m/\Lambda_{\rm ETC} \sim 10^{-3}\mbox{--}10^{-4}$,
where we took $M_\sigma = 500$~GeV.
This numerically agrees with 
that in the PCDC one~\cite{Bando:1986bg,Hashimoto:2010nw}. 

The gluon fusion process depends on the number $N_{TQ}$
of the techni-quarks and also the yukawa coupling, 
which is proportional to the number $N_D$ of the weak doublets 
for each TC index through the relation between $v^2$ and $m^2$.
The result $g_{\sigma ff}/g_{hff}^{\rm SM} \sim {\cal O}(1)$ for 
the one-family TC model near conformality with $M_\sigma \sim 500$~GeV 
implies that the production cross section of $\sigma$ is extremely enhanced.
This is noticeable, because the early stage of the LHC has the sensitivity to 
such a ``Higgs'' boson~\cite{ATLAS-higgs-sensitivity,CMS-higgs-sensitivity}.
On the other hand, in the models with smaller $N_{TQ}$ and/or $N_D$,
such a big enhancement is unlikely to occur.
In particular, the production cross section of $\sigma$ is suppressed
in the model having only one weak doublet of the techni-fermion 
and no techni-quark.

The branching ratios are also changed from the SM ones.
The main decay channel is expected to be the top pair, 
when the mass of $\sigma$ is above the threshold of $t\bar{t}$.


In this paper, we employed the ladder SD approach.
In the holographic WTC model, it is possible to calculate directly
the two point function $\Pi_\sigma(q)$, and 
thus $F_\sigma$ and also $M_\sigma$.
The analysis will be performed elsewhere~\cite{hWTC}.

In passing, we comment that 
several dynamical models predict existence of a (relatively) heavy Higgs,
which can be surveyed at the early stage of the LHC.
For example, the top condensate model with extra dimensions predicts 
successfully the top mass $m_t \simeq 175$~GeV~\cite{Hashimoto:2000uk}.
The Higgs mass is predicted as $m_H \sim $~200--300~GeV. 
The yukawa coupling is almost the same as the SM one.

The Higgs boson might reveal itself soon at the LHC.
The coming few years will be exciting.

\acknowledgments

The author thanks K.~Yamawaki for helpful discussions.

\end{document}